\documentclass[aps,preprint,prd,eqsecnum,nofootinbib,showpacs,12pt]{revtex4}

\usepackage{amssymb,latexsym} 
\usepackage{epsfig}
\usepackage{bm}

\newcommand{\be}{\begin{equation}} \newcommand{\ee}{\end{equation}}

\newcommand{\ba}{\begin{eqnarray}}
\newcommand{\ea}{\end{eqnarray}}

\begin{document}
\thispagestyle{empty} 

\date{\today}

\title{
{\vspace{-1.2em} \parbox{\hsize}{\hbox to \hsize 
{\hss \normalsize\rm IFUP-TH 2005/24, UPRF-2005-05}}} \\
Critical behavior in a non-local interface model
}

\author{Matteo Beccaria}%
\email{Matteo.Beccaria@le.infn.it}%
\affiliation{INFN, Sezione di Lecce, and
Dipartimento di Fisica dell'Universit\`a di Lecce,
Via Arnesano, ex Collegio Fiorini, I-73100 Lecce, Italy}

\author{Massimo Campostrini}%
\email{Massimo.Campostrini@df.unipi.it}%
\affiliation{INFN, Sezione di Pisa, and
Dipartimento di Fisica ``Enrico Fermi'' dell'Universit\`a di Pisa,
Largo Bruno Pontecorvo 3, I-56127 Pisa, Italy}

\author{Alessandra Feo}%
\email{feo@fis.unipr.it}%
\affiliation{
Dipartimento di Fisica, Universit\`a di Parma and INFN Gruppo Collegato di Parma,
Parco Area delle Scienze, 7/A, 43100 Parma, Italy}

\begin{abstract}
The Raise and Peel model is a recently proposed one-dimensional statistical model describing
a fluctuating interface. The evolution of the model follows from the competition between adsorption and desorption processes. 
The model is non-local due to the possible occurrence of avalanches.
At a special ratio of the adsorption-desorption rates the model is integrable and many
rigorous results are known. 
Off the critical point, the phase diagram and scaling properties are not known.
In this paper, we search for indications of phase transition studying the gap in the spectrum of the 
non-hermitian generator of the stochastic interface evolution.
We present results for the gap obtained  from exact diagonalization and from Monte Carlo estimates
derived from temporal correlations of various observables.
\end{abstract}

\pacs{05.10.-a, 05.40.-a, 05.10.yLn}

\maketitle


\section{Introduction}

The Raise and Peel model (RPM) \cite{deGier:2003dg} is a one-dimensional adsorption-desorption model of a fluctuating interface
defined on a discrete lattice with restricted solid-on-solid rules. 
The adsorption process raises the interface while in the desorption 
process the top layer of the interface evaporates (it ``peels'' the interface)
through non-local avalanches.
The RPM shows self organized criticality \cite{BTW,BV}.
The details of the dynamic will be presented in Subsection \ref{sec:definition}.

The continuum time limit of the model is characterized by a single coupling describing the ratio
$u$ of the interface adsorption/desorption rates. At $u=1$, the model describes a $c=0$ logarithmic conformal
field theory \cite{Kogan} with a dynamic critic exponent $z=1$.
It can also be analytically studied using its hidden Temperley-Lieb algebra \cite{deGier:2003dg,Rittenberg}.

At this value of the adsorption/desorption ratio,
many rigorous properties of the RPM are known about its equilibrium distribution, whose theoretical
properties are related to alternating sign matrices and combinatorics \cite{matrices,comb}.
In the continuum limit in the temporal direction the ratio $u$ can be any positive real number. 
In this paper we shall study the region $0<u \le 1$.

Away from the Temperley-Lieb rates, much less is known about the RPM.
In the preliminary analysis of Ref.~\cite{deGier:2003dg}, the RPM at $0<u<1$ has been proposed to have 
a critical point $u^*$, with $0<u^*<1$, separating two phases. The phase at $0<u<u^*$ is massive in the sense that the 
gap $E_1$ of the non-hermitian generator of the stochastic temporal evolution tends to a finite positive
value for large lattices. The phase in the region $u^* \le u \le 1$ is instead  a massless phase. 
The analysis of the behavior of $E_1(u, L)$ as a function of $u$ and the lattice size $L$ has been crucial in 
making this conjecture.

In this paper, we extend the analysis by pushing exact diagonalization up to the remarkable lattice 
length $L=26$. We also propose a Monte Carlo estimate of $E_1$ that we test up to $L=64$.

The analysis of the gap from these extended data provide information about the RPM phase diagram. 
Also, and most important, at the methodological level, it explores the finite size scaling properties of a non-local 
model discussing in a specific case the limits of the information that can be gained from this typical lattice sizes.

The outline of the paper is the following. In Section II we recall the definition of the RPM
and discuss the numerical simulation technique in the (time) continuum limit.
In Section III we recall the operatorial formulation of the model and 
summarize the conjectures about its phase diagram.
Results on the exact diagonalization on larger lattices are showed in Section IV. Section V 
gives an estimate of the first excited level of energy $E_1$ of this matrix while Section VI and VII
present our Monte Carlo procedure to simulate the model and the analysis of the
resulting data. Finally the Conclusions are 
summarized in Section VIII.

\section{The Raise and Peel model}
\label{sec:model}

\subsection{Definition of the model}
\label{sec:definition}

The RPM belongs to the class of  restricted solid-on-solid (RSOS) model. It is defined 
on a one dimensional lattice of size $L+1$, with even $L\in 2\mathbb{N}$. The sites are numbered from 0 to $L$, and 
at each site we assign a non negative height $h_i$ with RSOS constraints
\ba
| h_{i+1}-h_i | &=& 1, \\
h_0 = h_L &=& 0.
\ea
The space state ${\cal H}_L$ is discrete and its dimension is
\be
\dim {\cal H}_L = \frac{L!}{(L/2+1)(L/2)!^2}.
\ee

To assign a stochastic discrete dynamics on the space state ${\cal H}_L$, we first choose the two probabilities $u_a, u_d\in (0,1)$.
In the language of Ref.~\cite{deGier:2003dg}, they are associated to {\em adsorption} and {\em desorption} elementary 
processes. 
We then define the slope at site $i$ as $s_i = \frac{1}{2}(h_{i+1}-h_{i-1})$. The slope can only take values $s_i\in\{-1,0,1\}$.
\\ \\
At each temporal step, we make the following stochastic move:
\begin{enumerate}
\item choose at random a site $1\le i\le L-1$ with uniform probability $1/(L-1)$;
\item take the following actions (sub-moves)
   \begin{enumerate}
     \item[($\alpha)$:] if $s_i=0$ and $h_i>h_{i-1}$, do nothing;
     \item[($\beta$):] if $s_i=0$ and $h_i<h_{i-1}$, do $h_i\to h_i+2$ with probability $u_a$; 
     \item[($\gamma$):] if $s_i=1$, with probability $u_d$,
        find $j > i$ such that $h_j = h_i$ and $h_k > h_i$ 
        for all $i < k < j$ and make the transformation 
        $h_k\to h_k-2$ on all the internal sites $i < k < j$;
     \item[($\gamma'$):] if $s_i=-1$, with probability $u_d$, find 
        $j < i$ such that $h_j = h_i$ and $h_k > h_i$
        for all $j < k < i$ and make the transformation 
        $h_k\to h_k-2$ on all the internal sites $j < k < i$.
  \end{enumerate}
\end{enumerate}
It is easy to show that $N_\gamma = N_{\gamma'}$.

In the following we shall be interested in the time continuum limit of this stochastic chain. 
This is discussed in the next paragraph where we also give details of how the continuous stochastic dynamics
is practically simulated. 

\subsection{(Time) continuum limit and numerical simulation of its stochastic dynamics}
\label{sec:simulation}

The continuum limit in the temporal direction is taken at fixed $L$, by fixing the ratio $u\equiv u_a/u_d$
and sending $u_a, u_d\to 0$. The ratio $u$ can be any positive real number. However, in this work, we shall
consider mainly the region $0<u \le 1$, a restriction that we shall always assume if not said differently.
In the conclusions, we shall briefly mention the case $u>1$.

In the limit $u_d, u_a\to 0$, the lattice configuration is often
frozen for many Monte Carlo iterations.  Therefore, given a
configuration $s$ at time $t$, it would be very inefficient to perform
the simulation for each time step $t$, $t+1$, \dots; it is better to
compute the time interval $T$ during which $s$ stays unchanged, and
the new configuration $s'$ at time $t+T$.

We group the sites in four classes according to the above $\alpha$,
$\beta$, $\gamma$ and $\gamma'$ and denote by $N_\alpha$, $N_\beta$,
etc., the numbers of sites of a certain class.  We compute the total
transition probability $p=(u_a N_\beta + 2 u_d N_\gamma)/(L-1)$.  The
probability distribution of $T$ is proportional to $(1-p)^T$ and can
be sampled in the usual way by taking the logarithm of a uniform
random number.  Finally, we extract the site class $\beta$, $\gamma$
or $\gamma'$ with probability proportional to $u_a N_\beta$, $u_d
N_\gamma$, $u_d N_\gamma$, respectively, and pick the site $i$ in the
class with uniform probability.

\section{Operatorial description of the RPM}

It is convenient to discuss the (continuum time limit of the) RPM and its dynamics with a Hilbert space formalism
where we can easily rephrase all statements concerning the underlying stochastic process.
We associate the vector $|s\rangle$  with each configuration $s$ of the RPM. 
We denote by ${\cal H}$ the finite dimensional complex vector space having these states as an
orthonormal basis
\be
\langle s | s'\rangle = \delta_{ss'}.
\ee
The subset ${\cal H}_S\subset{\cal H}$ is the set of {\em stochastic vectors} of the form 
\be
\sum_i p_i |s_i\rangle,\qquad 0\le p_i\le 1,\qquad \sum_i p_i = 1.
\ee
They can be used to describe the state probability distribution at a certain time.

We assign a special name to the state 
\be
|\sigma\rangle = \sum_s |s\rangle.
\ee
It can be used to implement the sum over all states by taking scalar products.

The evolution of the RPM is described by means of a non hermitian rate matrix $H$.
The specific matrix elements of $H$ for $L=6$ can be found in Ref.~\cite{deGier:2003dg}.

An important property of $H$ is the conservation of probability expressed by the relation
\be
\langle\sigma|H = 0.
\ee
Given $H$, we define the evolution operator $U(t) = e^{-tH}$. It has the following meaning
\be
U(t)_{s's} = \langle s' | U(t) | s\rangle = \mbox{Prob}(s, t_0\to s', t_0+t) .
\ee
In particular, if $|\psi\rangle\in{\cal H}_S$ is a stochastic vector describing a statistical ensemble of states
at time $t_0$, then the stochastic vector at time $t_0+t$ is given by 
\be
U(t) |\psi\rangle \in {\cal H}_S.
\ee

The eigenvalues of $H$ can be complex, and we sort them by increasing real part:
\be
H|\psi_n\rangle = E_n |\psi_n\rangle,\qquad 0=E_0 < \mbox{Re}\ E_1 < \mbox{Re}\ E_2 < \dots ,
\ee
It is convenient to introduce the expansion of the eigenvectors in terms of the single configuration 
basis
\be
|\psi_n\rangle = \sum_s \psi_{n,s} |s\rangle .
\ee
The eigenvectors $|\psi_n\rangle$ are not necessarily orthogonal. In the following we shall always assume
that they are a basis of the state space. 
The condition $\mbox{Re} E_i\ge 0$ expresses the convergence property of the stochastic process
described by $H$. Due to this condition we have 
\be
\lim_{t\to\infty} U(t)_{s's} = \psi_{0,s'},
\ee
independently on the initial state $s$. In other words, every statistical ensemble converges
asymptotically to the equilibrium distribution specified by the components of the lowest
eigenvector. Due to the meaning of $|\psi_0\rangle$, it is convenient to choose the normalization of 
$|\psi_0\rangle$ by requiring that it belongs to ${\cal H}_S$
\be
\sum_s\psi_{0,s} = 1, \qquad\mbox{or}\qquad \langle\sigma | \psi_0\rangle = 1 ,
\ee
in agreement with its probabilistic interpretation.

\subsection{The RPM away from the Temperley-Lieb point}

The special point $u=1$ defines the so-called Temperley-Lieb (TL) rates. At this value of the 
adsorption/desorption ratio the RPM can be connected to the dense O(1) or TL loop model~\cite{TL}
and many rigorous properties are known about its equilibrium distribution.

In particular, at $u=1$ the model is described by $c=0$ logarithmic conformal field theory (LCFT)
with a dynamic critical exponent $z=1$.

Away from the Temperley-Lieb rates, much less is known about the RPM. The aim of this paper is
precisely the investigation of the phase structure of the model with respect to the rate variable $u$.

Following the analysis of Ref.~\cite{deGier:2003dg}, the existence of possible critical points for $u<1$
can be studied by computing the first excited level $E_1$ as a function of the rate $u$ and the lattice
size $L$. From exact diagonalization up to $L=16$ and finite size
scaling analysis, the authors of Ref.~\cite{deGier:2003dg} have 
claimed that the RPM exists in two phases when $u<1$. The first phase is called massive and is 
realized for $0<u<u^*$. The second phase, for $u^*\le u\le 1$ is a massless phase.

As we mentioned in the Introduction, the naming refers to the different behavior of $E_1(u, L)$ at large $L$. In a massive phase
$E_1$ tends to a positive constant as $L\to\infty$. In a massless phase, it tends to zero.
At $u=1$, we know rigorously that $L E_1$ tends to a positive constant. 

The finite size analysis of Ref.~\cite{deGier:2003dg} studies the crossing of $LE_1(u, L)$ between the curves at the two 
values $L$ and $L+2$ with $L\le 16$. Two crossings are observed. The smaller one appears to converge to some value $u^*$
around 0.5 as $L$ increases. The larger crossing is less stable and it possible to conjecture that it tends to 1
at infinite $L$. Due to the small lattice sizes, these conclusions are quite preliminary and deserve 
a more detailed investigation.

In the next Sections, we shall first discuss the outcome of an improved analysis based on 
exact diagonalization up to the size $L=26$. Then, we shall illustrate a Monte Carlo method
to compute $E_1$ and numerical data up to $L=64$. These are less precise than exact diagonalization points,
but can complement them, as we shall illustrate.

To conclude this Section, we like to add a general comment. In principle, it is possible to study the critical
behavior of the RPM by alternative methods, in particular by studying the size and $u$ dependence
of statistical averages over very large lattices. In our analysis we concentrated on the (difficult)
measure of $E_1$ with the precise aim of extending the above analysis.

\section{Exact Diagonalization on Large Lattices}
\label{sec:ED}

\begin{center}
\begin{figure}[htb]
\epsfig{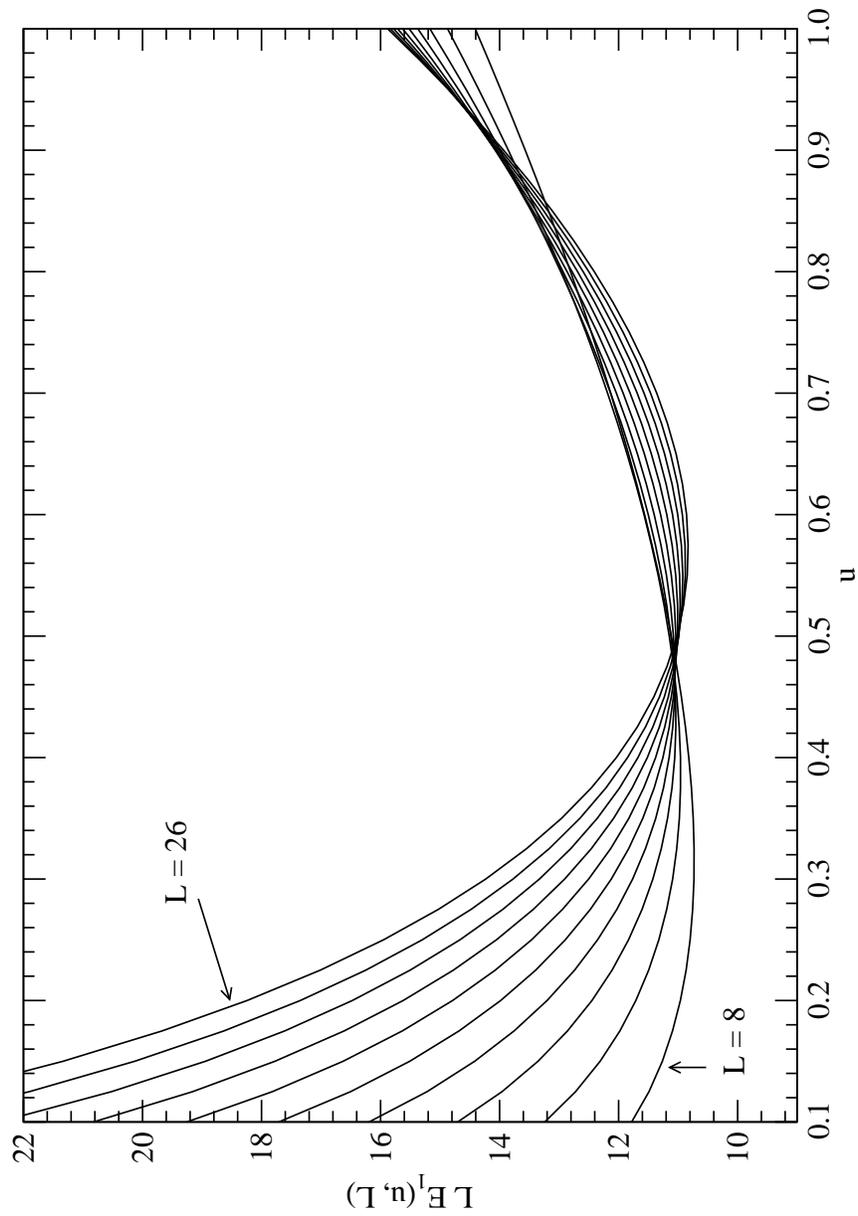}
\caption{
Exact diagonalization data for the function $LE_1(u, L)$. The various lines shows the results
at sizes $L=8, 10, \dots, 24, 26$.
}
\label{fig:ED}
\end{figure}
\end{center}

\begin{center}
\begin{figure}[htb]
\epsfig{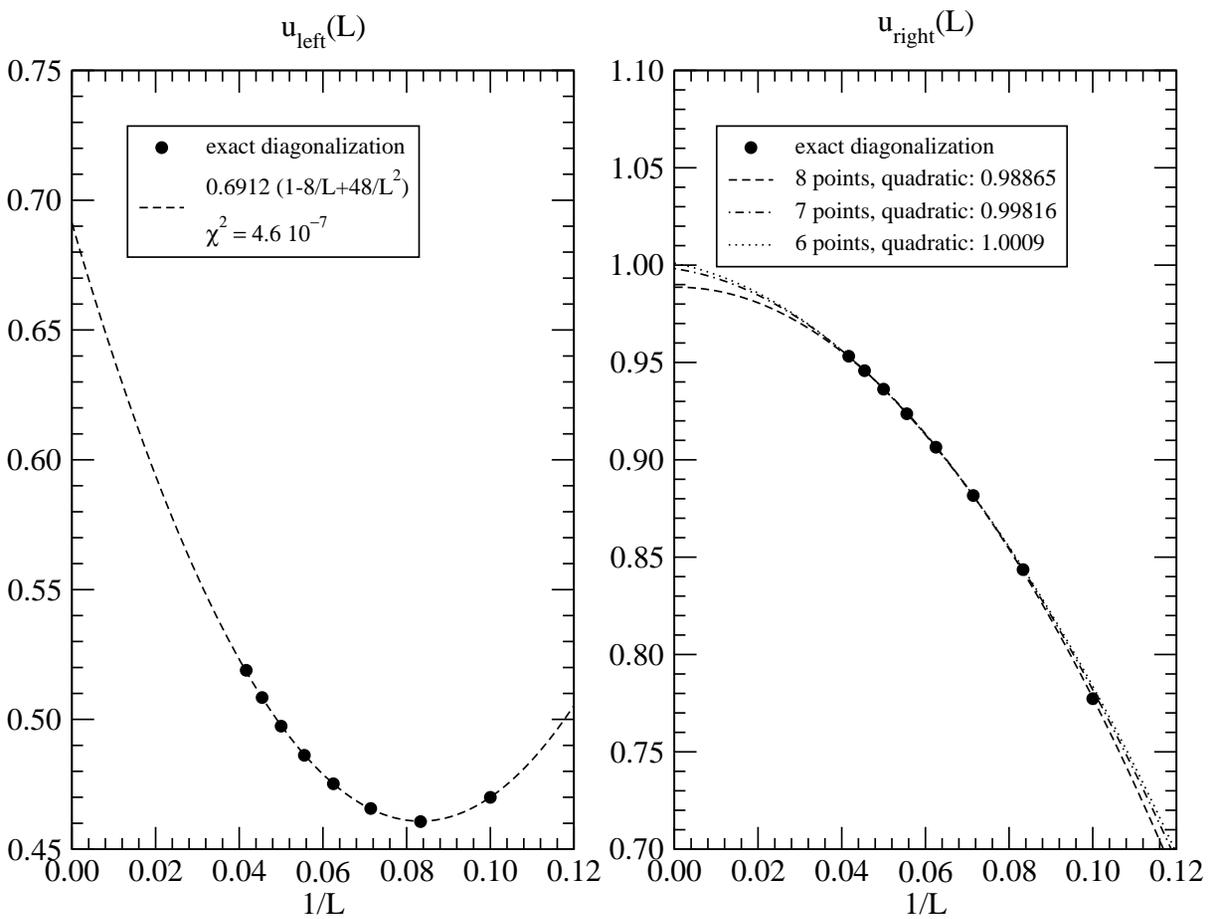}
\caption{
Exact diagonalization data for the two crossings $u_{\rm left, \ right}(L)$ discussed in Sec.~(\ref{sec:ED}).
}
\label{fig:cross}
\end{figure}
\end{center}

We have pushed the exact diagonalization of the RPM Hamiltonian up to the size $L=26$, using the software
package ARPACK~\cite{ARPACK}.
A plot of $LE_1(u, L))$ as a function of $u$ for the various considered $L$ is shown in Fig.~(\ref{fig:ED}).
{}From that figure, we can see the existence of two crossings between the curves at $L$ and $L+2$. We shall denote
them as $u_{\rm left}(L)$ and $u_{\rm right}(L)$ with $u_{\rm left}(L) < u_{\rm right}(L)$. 
We want to compute the extrapolations
\be
u^*_{\rm left, \ right} = \lim_{L\to\infty} u_{\rm left, \ right}(L).
\ee
In Fig.~(\ref{fig:cross}) we show the crossings  $u_{\rm left}(L)$, $u_{\rm right}(L)$ as functions of $1/L$
to test convergence. The left part of the figure shows $u_{\rm left}(L)$. Data are accurately fit 
by the simple function 
\be
0.6912\ (1-8/L+48/L^2).
\ee
The integer coefficients are empirical, but the quality of the fit
is remarkable with a $\chi^2=4.6\cdot10^{-7}$.\footnote{Here and in the
following, we denote by $\chi^2$ the sum of the squares of the
differences between the data points and the values of the fitting
function.} Our estimated extrapolation $u^*_{\rm left} = 0.6912$
suggests the existence of a critical point in infinite volume that
separate different phases of the model.

Of course, without further analytical insight, it is also perfectly
possible that the above extrapolation is fake and that we are working
on a lattice which is too small to extrapolate reliably to the true
infinite size limit.  We leave this kind of comments to the
conclusions.

The analysis is similar for $u_{\rm right}(L)$ shown in the right part of the figure. In this case, 
we also tried a polynomial in $1/L$, but did not find a simple expression for the coefficients
of the various powers of $1/L$. The figure shows three curves obtained by fitting the first 8, 7, 6
data points with the highest $L$. As data points with smaller $L$ are discarded, the extrapolated estimate
for $u_{\rm right}^*$ moves closer and closer to the special value $1$. Thus our data support the 
conjecture $u^*_{\rm right} = 1$ with no additional non trivial critical point.
The convergence in $L$ is quite slow; this is not surprising,
given the non-local nature of the dynamics.

At size $L=26$, the state space dimension is 742900. For computational reasons, it is not sensible to 
push the exact diagonalization beyond this point. Indeed, the convergence rate can be judged from the above figures and is 
not dramatically fast. 

For this reason, it is convenient to resort to Monte Carlo methods. In the next Sections, we shall 
first describe our theoretical approach, and then discuss the actual measurements of $E_1$ on larger lattices.

\section{Stochastic averages and estimate of $E_1(u, L)$}
\label{sec:operator}

The RPM is a Markov chain where the non hermitian Hamiltonian plays the role of a 
rate matrix. Our analysis of the phase structure of the RPM is based on the calculation
of the first excited level $E_1$ of this matrix. In this Section, we recall the basic 
formalism that allows to extract $E_1$ from the time evolution of certain stochastic averages
evaluated over the chain.

\subsection{Expansion of vectors}

We assumed that the eigenvectors of $H$ are a basis of ${\cal H}$. Hence, 
any state $|\phi\rangle$ can be expanded in terms of the eigenvectors of $H$
\be
|\phi\rangle = \sum_{n\ge 0} c_n |\psi_n\rangle .
\ee
The coefficient $c_0$ can be determined. We apply $U(t)$ to the above expansion and take the $t\to +\infty$ limit.
Since $E_n>0$ for $n>0$, we find 
\be
\lim_{t\to+\infty} U(t)|\phi\rangle = c_0 |\psi_0\rangle .
\ee
Taking the scalar product with $|\sigma\rangle$ and exploiting $\langle\sigma|U = \langle\sigma | $ as well as
the chosen normalization of $|\psi_0\rangle$ we conclude that     $c_0=\langle\sigma|\phi\rangle$, i.e., 
\be
|\phi\rangle = \langle\sigma|\phi\rangle |\psi_0\rangle  + \sum_{n\ge 1} c_n |\psi_n\rangle . 
\ee

\subsection{Time averages and their operator expression}

Let us consider various functions $\{A_i(s)\}_{1\le i\le N}$ of the RPM configurations. The stochastic 
$N$-point averages times $\{t_i\}$ 
starting from a given configuration $s_0$ will be denoted by 
\be
\mbox{\bf E}_{s_0}\left[\prod_{i=1}^N A_{i}(s_{t_i})\right],
\ee
where the expectation is over all realizations of the stochastic process $\{s_t\}_{t\ge 0}$
with assigned initial condition. The equilibrium averages and correlations are independent on the
initial state and are defined as
\ba
\mbox{\bf E}\left[A\right] &=& \lim_{t\to+\infty}\ \mbox{\bf E}_{\rm any}\left[ A(t)\right], \\
\mbox{\bf E}\left[A(0) B(\tau)\right] &=& \lim_{t\to+\infty}\ \mbox{\bf E}_{\rm any}\left[ A(t)B(t+\tau)\right],\\
&\cdots&
\ea

As a preparation to the calculation of stochastic time averages, we associate an operator
to each function $A(s)$ of the RPM configuration $s$. The rule is trivial. We associate to $A(s)$
the operator defined by the following diagonal matrix elements in the $|s\rangle$ basis
\be
\langle s | A | s'\rangle = A(s) \delta_{ss'}.
\ee

Consider now the stochastic average of a single $A$ at time $t$, starting from the fixed configuration  $s_0$.
By definition, we can write
\be
\mbox{\bf E}_{s_0}\left[A(s_t)\right] = \langle \sigma |A\ U(t)| s_0\rangle .
\ee
{}From the above expansion we have
\be
|s_0\rangle = |\psi_0\rangle  + \sum_{n\ge 1} c_n |\psi_n\rangle 
\ee
with certain coefficients $c_n$. Thus, 
\be
\mbox{\bf E}_{s_0}\left[A(s_t)\right] = \langle \sigma|A|\psi_0\rangle + \sum_{n\ge 1} c_n e^{-tE_n} \langle\sigma|A| \psi_n\rangle .
\ee
Of course, the asymptotic value is precisely the equilibrium average 
\be
\mbox{\bf E}\left[A\right] = \langle \sigma | A | \psi_0\rangle = \sum_s A_s \psi_{0,s} .
\ee

Another example is the self correlation of $A$ with lag $\tau$. By definition, this is 
\ba
\mbox{\bf E}\left[A(0) A(\tau)\right]  &=& \sum_{ss'} A_{s'}\mbox{prob}(s\to s') A_s\psi_{0,s} \\
&=& \sum_{ss'} A_{s'}U(\tau)_{s's} A_s\psi_{0,s} = \langle\sigma| A U(\tau) A |\psi_0\rangle \nonumber
\ea
Again, we have the expansion 
\be
A|\psi_0\rangle = b_0 |\psi_0\rangle + \sum_{n\ge 1} b_n |\psi_n\rangle
\ee
for certain coefficients $b_n$. Then, the large $\tau$ behavior of the connected correlation is given by 
\be
\mbox{\bf E}\left[A(0) A(\tau)\right] - \mbox{\bf E}\left[A\right]^2 = \sum_{n\ge 1} c_n \langle\sigma | A|\psi_0\rangle e^{-E_n\tau}
\ee

\section{Monte Carlo simulation and estimate of $E_1$}

The Monte Carlo procedure consists of the following steps.

A thermalized configuration $\tilde{s}$ is generated by a
straightforward implementation of the update procedure described in
Subsection~\ref{sec:simulation}.

A set of $b$ {\em branches} is generated by running $b$ times the
update procedure for a time duration $\tau$, each branch starting from
$\tilde{s}$, with a different sequence of random numbers.  Along each
branch, the values of the observables $A_i(t)$ are averaged over time
intervals of size $\Delta t$; the resulting values are used to
estimate the expectation value of $A_i(r\,\Delta t)$ with initial
conditions $s(0) = \tilde{s}$: 
\be
A_{i,r} = \bar A_i(r\,\Delta t) = 
{1\over\Delta t}\int_{(r-1/2)\Delta t}^{(r+1/2)\Delta t}
\langle \sigma |A\ U(t)| \tilde{s}\rangle.
\ee
We selected the following set of observables:
\be
A_0 = h_{L/2},\quad A_1 = \sum_{i=1}^{L-1} h_i, \quad
   A_{d+2} = \sum_{i=1}^{L-1-d} h_i h_{i+d},\ 0\le d\le14.
\ee
We take as value and statistical error of $A_{i,r}$ the average and
standard deviation over the branches at fixed time interval.  We
typically choose $b=10^8$.

The resulting data for $A_{i,r}$ are fitted to the two-exponential
formula
\be
A_{i,r} = a^{(i)}_0 + a^{(i)}_1 \exp(-\mu^{(i)}_1\,r\,\Delta t)
+ a^{(i)}_2 \exp(-\mu^{(i)}_2\,r\,\Delta t),
\ee 
with $\mu_1<\mu_2$, for $r\,\Delta t>t_{\rm min}$.  The value and
error of $\mu_1$ are estimated checking the stability of the fit vs.\ 
$t_{\rm min}$ and the consistency of the best-fit values of $\mu_1$
between different observables.

The procedure is repeated starting the branching process from several
statistically-independent configurations $\tilde{s}$ and the resulting
values of $\mu_1$ are analyzed.  If a configuration gives a value
inconsistent with the others, it is removed.  The value and error of
$\mu_1$ are estimated from the remaining configuration by a standard
statistical analysis; the error is multiplied by the scale factor 
$s = \sqrt{\chi^2/n_{\rm d.o.f}}$ if $s>1$, i.e., if the values are
not fully consistent within errors; we consider $s^2>6$ a warning sign
of possible troubles.  The resulting value and error of $\mu_1$ is
directly an estimate of $E_1$.

\begin{center}
\begin{figure}[htb]
\epsfig{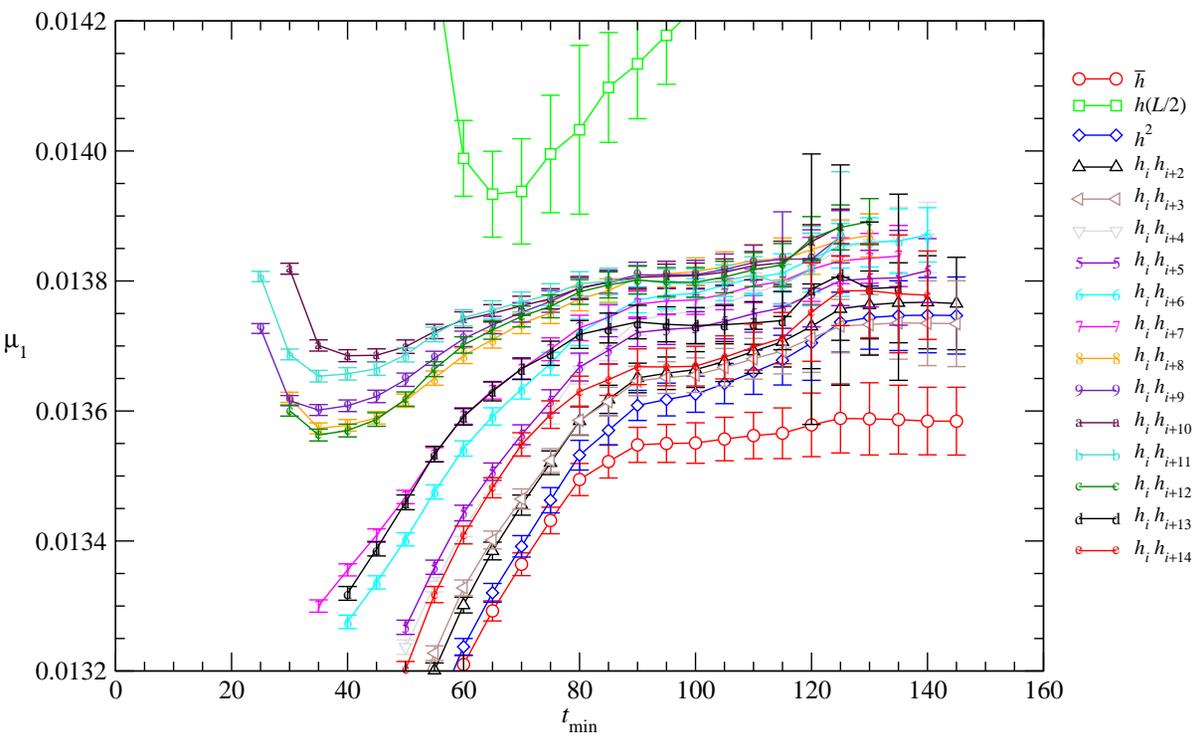}
\caption{
$\mu_1$ vs.\ $t_{\rm min}$ for $u=0.9$ and $L=32$.
Discarding $A_0$ (squares) and $A_1$ (circles),
we estimate $\mu_1=0.0135 \div 0.0140$.
}
\label{fig:mu1vst}
\end{figure}
\end{center}

A typical example is shown in Fig.~\ref{fig:mu1vst}.
Some branch sets give extremely unstable results and are discarded
from the analysis; it is likely that their initial configurations
happen to have a very small superposition on $|\psi_0\rangle$. 
It turns out that $A_0$ and $A_1$ often give unstable fit results and
are not consistent with each other and with other $A_i$; therefore we
exclude them from the analysis.

\section{Data analysis and Discussion}
\label{sec:discussion}

We study the behavior of $E_1$ vs.\ $L$ at fixed $u$.  We fit the
exact data for $8\le L\le26$ to one of the forms
\be 
L E_1(L) = c + a_1 L^{-\nu} + a_2 L^{-2\nu} 
\ee
(for the massless phase) and
\be 
E_1(L) = c + a_1 \exp(-\mu_1 L) + a_2 \exp(-\mu_2 L) 
\ee
(for the massive phase).  We discriminate the phase by the value of
the $\chi^2$ of the fit and by consistency of the fit, extrapolated to
large $L$, with Monte Carlo data.

\begin{center}
\begin{figure}[htb]
\epsfig{width=12cm,file=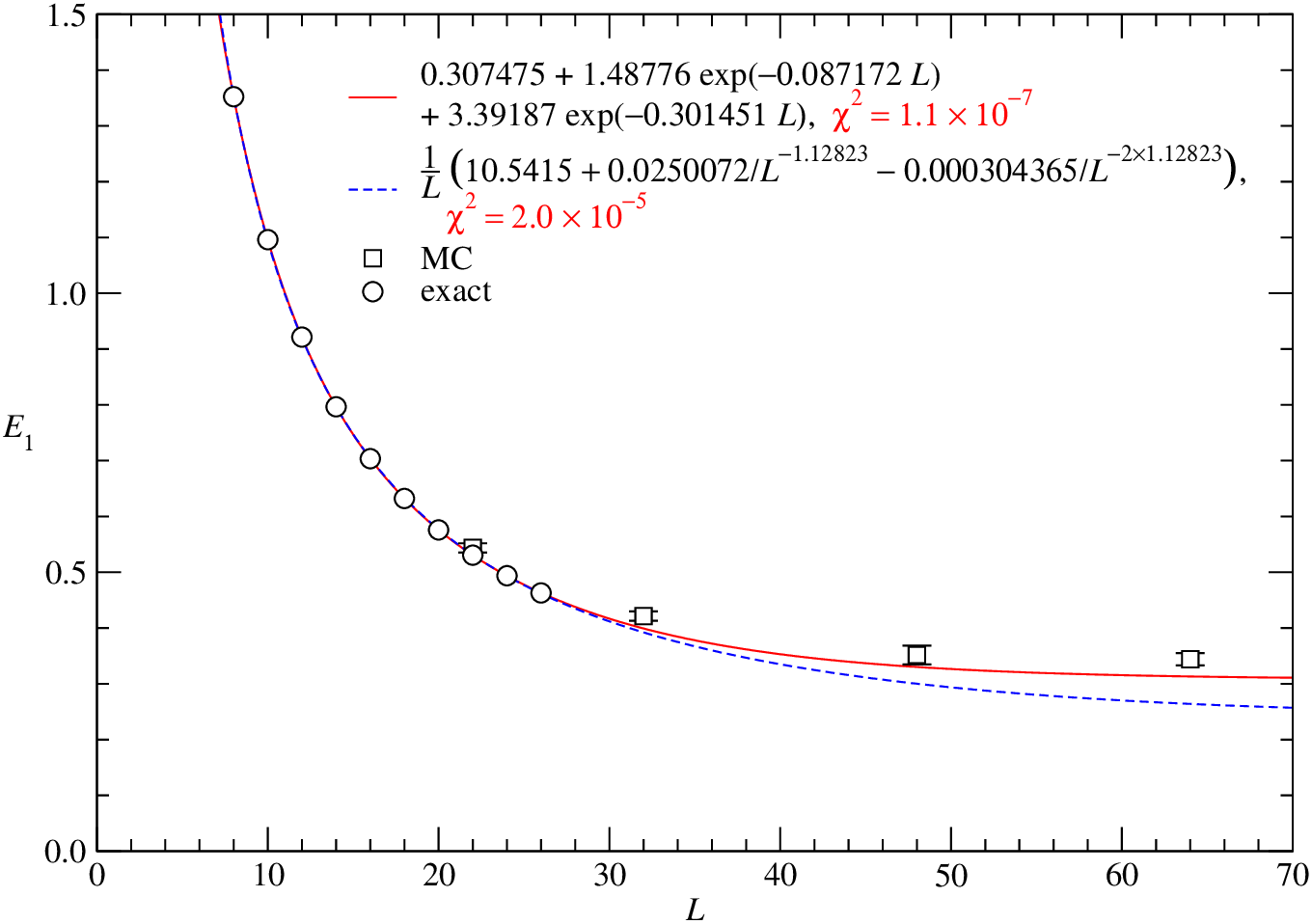}
\caption{
$E_1$ vs.\ $L$ for $u=0.4$.  Both $\chi^2$ and MC data clearly favor
the massive phase.
}
\label{fig:u40}
\end{figure}
\end{center}

\begin{center}
\begin{figure}[htb]
\epsfig{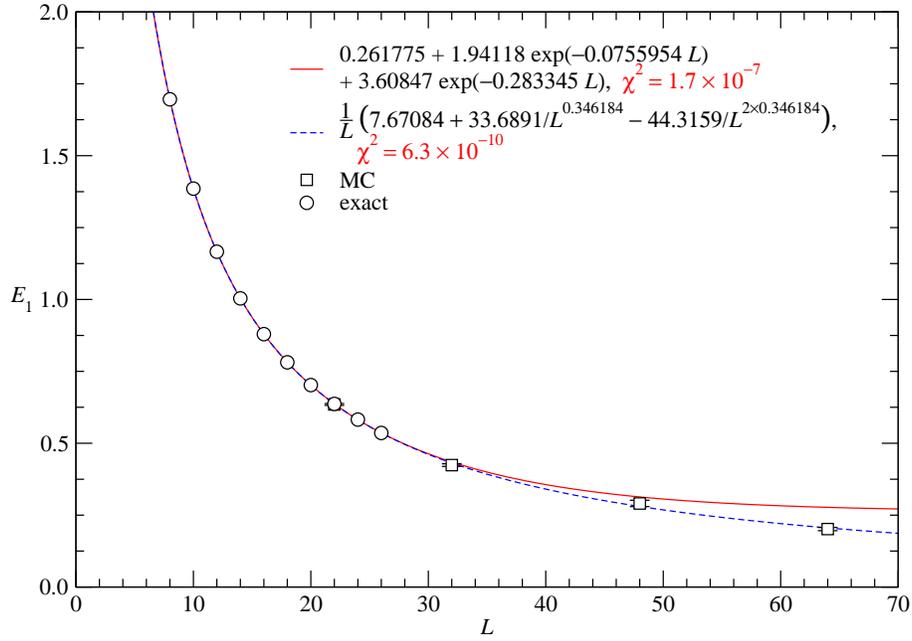}
\caption{
$E_1$ vs.\ $L$ for $u=0.9$.  Both $\chi^2$ and MC data clearly favor
the massless phase.
}
\label{fig:u90}
\end{figure}
\end{center}

Two typical examples are shown in Figs.~\ref{fig:u40} and
\ref{fig:u90} where we show the results of the analysis at the two points $u=0.4$, and $u=0.9$.
The former value is clearly in the massive phase, the latter in the massless one. This is 
consistent with Fig.~(\ref{fig:cross}) at all considered $u$ values, {\em i.e.} from 0.1 to 1.0
with 0.1 spacing. Of course, the discrimination in the critical region $u\simeq 0.7$ is 
more difficult and less clear. 
In the massless phase, the determination of $\nu$ is quite imprecise;
it favors values $\nu\sim1/2$, although, i.e., for $u=1$ it is also
compatible with $\nu=1$. (In the massive phase, the ``wrong''
exponential fit often gives a negative $\nu$.)

\begin{center}
\begin{figure}[htb]
\epsfig{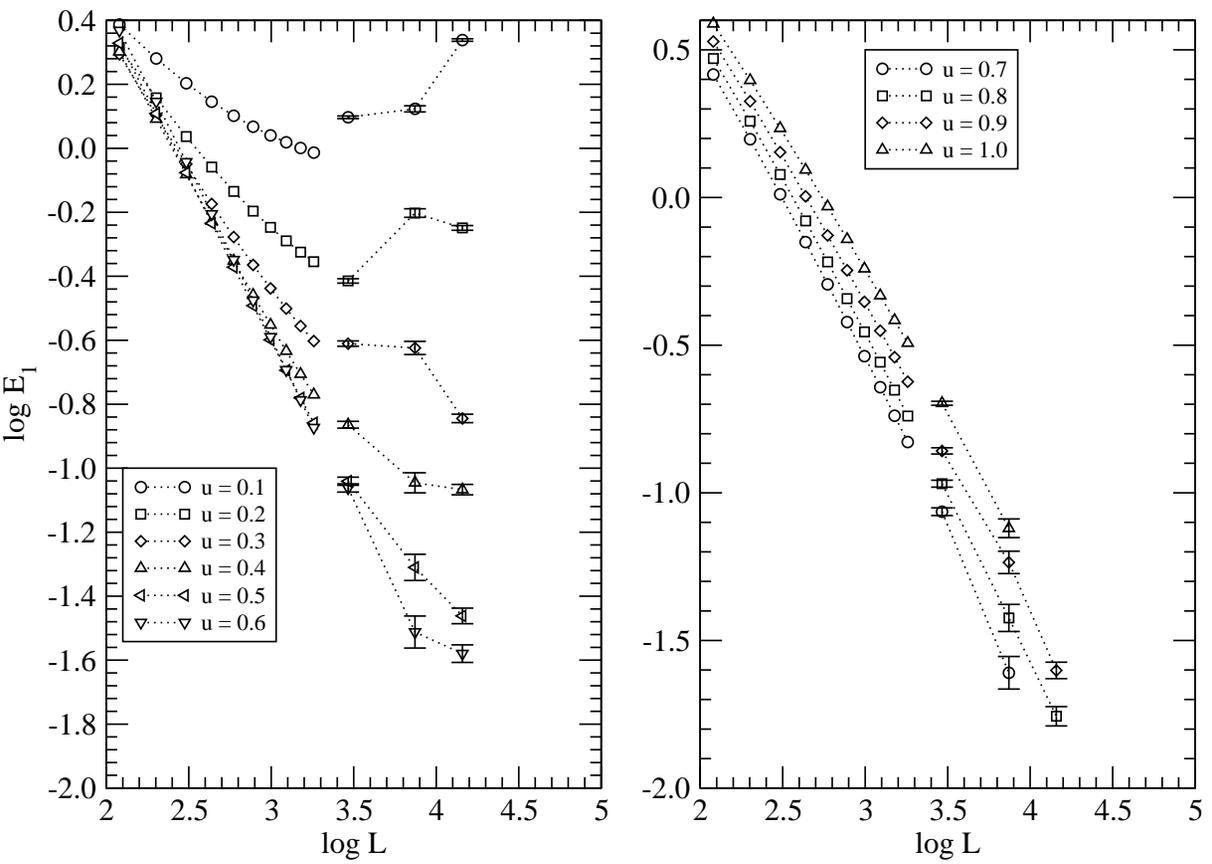}
\caption{
Exact Diagonalization and Monte Carlo data for $E_1(u, L)$ vs. $L$ in log-log scale at
$u = 0.1-1.0$ in steps $\Delta u=0.1$.
}
\label{fig:summary}
\end{figure}
\end{center}

A combination of the exact diagonalization and Monte Carlo data is summarized in Fig.~(\ref{fig:summary}).
In each panel we have shown in log-log scale the ED data and the three points from Monte Carlo ($L=32$, $48$, and $64$)
versus $L$.

The left panel contains data at $u=0.1-0.6$. The ED data at 0.1-0.3 bend in the up direction suggesting 
convergence of $E_1$ to a finite positive limit as $L\to\infty$. This conclusion is strongly supported by 
MC data. The bending is not visible at $u = 0.4$, $0.5$, and $0.6$, but again MC data suggest the same asymptotic
behavior. 

The right panel of Fig.~(\ref{fig:summary})  shows the remaining values $u=0.7-1.0$. In this case, the ED data 
bend in the down direction in agreement with an algebraic decay of $E_1$ with the lattice size. 
This is confirmed by the MC points that extend smoothly this trend. In other words, at the explored 
lattice sizes, there are no signal for a plateau in $E_1$. 

We also find remarkable that the special point $u^*\simeq 0.69$ singled out by the exact diagonalization analysis 
remains a sort of separator point between two qualitatively different behaviors also after the inclusion of 
the Monte Carlo data points.

\section{Summary and Discussion}

In summary, we have investigated the RPM away from its integrable point. 
First, we have extended the available results from exact diagonalization of the non local RPM Hamiltonian.
As a further step, we have exploited Monte Carlo methods to obtain spectral information on the Hamiltonian
from the temporal decay of stochastic averages of various observables.

As suggested in Ref.~\cite{deGier:2003dg}, we have found evidence for a drastic change of behavior
of the system at the intermediate adsorption/desorption rate ratio $u^*\simeq 0.69$. The analysis of data 
on lattices up to 64 sites long has confirmed rather clearly that the RPM is in a massive phase for $u<u^*$
where a plateau can be seen in the plot of $E_1$ vs. $L$. The region $u^*<u\le 1$, where 
our data show a steady decrease in $E_1$, is more difficult to analyze. Monte Carlo simulations
cannot exclude that $E_1(u, L)$ reaches a positive asymptotic value at a typical lattice size growing 
rapidly as $u\to 1$.
To exclude such a scenario it would be necessary to investigate the phase diagram of the RPM with other techniques
able to reach very large lattices. We believe that a determination of $E_1$ from Monte Carlo simulations 
is not feasible on lattices much larger than the length we explored. However, it would be possible
to measure static averages of typical non local quantities, like cluster distributions in the 
interface configurations, with good accuracy on quite larger lattices. Preliminary results
along this line seem to suggest the existence of a single massive phase for all $u<1$~\footnote{V. Rittenberg, private
communication}. If such a scenario were confirmed, the Raise and Peel model would be an interesting example
of very slow convergence to criticality in the infinite volume limit, presumably related to the non-locality
of the model.
We emphasize again that the RPM is non-local at generic $u$ because of the occurrence of avalanches. 
This means that the Hamiltonian $H$ does not describe a lattice model with local interaction and it is not
straightforward to apply field theory techniques. Only at the TL point $u=1$, the hidden loop algebra permits
a straightforward and clean  analytical investigation. Off integrability, some better degrees of freedom 
must be identified for an effective description of the avalanche dynamics.  \\

Apart from these speculations, we believe that our proposal for a direct measure of $E_1$ is valuable
since the gap is often an important quantity to be measured on critical discrete models.
In principle indeed, the strategy described in this paper could be applied to more involved versions of the 
RPM. In particular, it would be interesting to analyze the role of boundary conditions as 
discussed in Ref.~\cite{Pyatov}.

\acknowledgments

We would like to thank Vladimir Rittenberg for enlightening discussions. 

The calculations have been done on a PC Cluster at the Department of Physics of the University of Parma.

\newpage

\newpage

\newpage

\newpage

\newpage

\end{document}